\documentclass{IEEEtran}
\usepackage{amssymb,amsfonts}

\let\phi\varphi
\def\w{w}
\newtheorem{theorem}{Theorem}[section]

\newtheorem{lemma}[theorem]{Lemma}
\newtheorem{DDDefinition}[theorem]{Definition}
\def\enddefinition{\end{DDDefinition}\egroup\medbreak}
\makeatletter
\def\definition{\bgroup\def\@begintheorem##1##2{\trivlist
  \item[\hskip\labelsep{\bfseries ##1\ ##2}]}\begin{DDDefinition}}
\makeatother

\begin{document}
\author{L\'aszl\'o Csirmaz\thanks{Central European University and University
of Debrecen. Research was partially supported by grant NKTH
OM-00289/2008 and the ``Lend\"ulet'' project.}
and G\'abor
Tardos\thanks{School of Computing Science, Simon Fraser University, Burnaby,
  BC and R\'enyi Institute of Mathematics, Budapest Research was
  partially supported by NSERC Discovery grant, the Hungarian OTKA grants
  T-046234, AT-048826, NK-62321 and the ``Lend\"ulet'' project.}}
\date{}
\title{\bf Optimal information rate of secret sharing schemes on trees}

\maketitle

\begin{abstract}
The information rate for an access structure is the reciprocal of the
load of the optimal secret sharing scheme for this
structure. We determine this value for all trees: it is $(2-1/c)^{-1}$, where
$c$ is the size of the largest core of the tree. A subset of the vertices of a
tree is a {\it core} if it induces a connected subgraph and for each vertex in
the subset one finds a neighbor outside the subset. Our result follows from a
lower and an upper bound on the information rate that applies for any graph
and happen to coincide for trees because of a correspondence between the size of
the largest core and a quantity related to a fractional cover of the tree
with stars.

\medskip
\noindent{\bf Keywords. } Secret sharing scheme; information rate; graph; 
fractional packing and cover; entropy method.

\noindent{\bf AMS classification numbers. } 05B40, 05C85, 94A60, 94A62,
94A17
\end{abstract}

\section{Introduction}\label{sec:intro}
Secret sharing schemes has been investigated in several papers, for an
extended bibliography see \cite{stinson-wei}. Such a scheme with $n$
participants is a joint
distribution of $n+1$ discreet random variables, one called the {\em secret},
the rest being the {\em shares} of the participants. An {\em access structure}
designates certain subsets of the participants as {\em qualified} leaving the
rest of the subsets {\em unqualified}. A secret
sharing scheme for an access access structure has to satisfy that
one can recover the secret with probability 1 from the shares
of any qualified subset of the participants but the secret should be
statistically
independent from the collection of shares belonging to an unqualified subset.

In this paper we deal with access structures based on graphs. The scheme is
{\em based on the graph $G$} if the participants are the vertices, and
unqualified subsets are the independent sets. This makes the endpoints of the
edges the minimal qualified subsets. We simply call a secret sharing scheme
for the access structure based on a graph $G$ a secret sharing scheme on $G$.

The {\it load} of a scheme is measured by the amount of 
information the most heavily loaded participant must remember for each bit
in the secret. Formally, this is $\max_i(H(S_i))/H(\xi)$, where $S_i$ is the
share of participant $i$, $\xi$ is the secret and $H$ denotes entropy. We
assume $H(\xi)>0$. For a graph $G$ the {\it information complexity of $G$}, 
denoted as $\sigma(G)$, is the infimum of the loads of all
secret sharing schemes on $G$. The
{\it information rate}, usually denoted as $\rho(G)$, is simply
$\rho(G)=1/\sigma(G)$, the inverse of
this value. The notation $\sigma(G)$ for the complexity of the scheme was
introduced in \cite{farre-padro}.
The information rate of graphs has been investigated in several papers, see
\cite{dijk} for the rate of graphs with at most six
vertices and also (among other works) \cite{blundo2, blundo3, blundo4,
brickell1, brickell2, capocelli, stinson}.

In \cite{stinson} Stinson describes a general secret sharing construction, 
which, 
when applied to graphs, gives the upper bound $(d+1)/2$ for the complexity of
graphs with maximum degree $d$. Blundo et al.\ in \cite{blundo2} constructed
an infinite family of graphs for each $d$  for which Stinson's bound is tight. 
The $d=2$ case is fully settled in \cite{blundo3}: the information complexity
of paths and cycles is $3/2$ except for $P_2$, $P_3$,
$C_3$ and $C_4$, when it is $1$. The information complexity of the
$d$-regular $d$-dimensional hypercube is exactly $d/2$,
see \cite{csirmaz1}. Our paper is the first one which determines the
information complexity and information rate of graphs in a large and natural
family, namely, for trees.

\medskip

To state our result we need the notions of {\it core} and {\em star cover
rate} of an arbitrary graph.

\begin{definition}\label{def:core}
We call a subset $X$ of the vertices of a graph $G$ a {\em core} of $G$ if
it induces a connected subgraph and one
can find a neighbor $x'\notin X$ of each $x\in X$ such that $x$ is the only
neighbor of $x'$ among the vertices in $X$ and $\{x'\mid x\in X\}$ is an
independent set.

A {\em fractional star packing} in a graph $G$ is a collection of star
subgraphs of $G$, each with an associated positive weight. The weight of a
vertex or an edge in a fractional star packing is the total weights associated
to stars containing that vertex or edge, respectively. The {\em star cover
rate} of $G$ is the infimum (minimum) of the maximal vertex weights among all
fractional star packings with each edge having weight at least 1.

If the weights in a fractional star packing are integral we speak of {\em star
packing} and we say a vertex or edge is {\em covered $k$ times} if its weight
is $k$.
\end{definition}

\noindent
Notice that when $G$ is a tree a subset $X$ of its vertices is a core if it
induces a connected subgraph and each $x\in X$ has a neighbor outside $X$.

\begin{theorem}\label{thm:main}
Let $G$ be a graph, let $c=c(G)$ be the maximum size of a core of $G$ and let
$s=s(G)$ be the star cover rate of $G$. For the information complexity
$\sigma(G)$ of $G$ we have
$$2-1/c\le\sigma(G)\le s.$$
\end{theorem}

Note that the second inequality of this theorem comes from Stinson
\cite{stinson}. We state it here for completeness. Both the lower and the
upper bounds are often useful, but they are not tight in general. The
graph $\Delta$ depicted in Figure~1 has only one vertex cores, its
information complexity is $3/2$ and its star cover rate is $5/3$. Thus we have
strict inequalities in
$$
    2-1/c(\Delta) < \sigma(\Delta) < s(\Delta).
$$

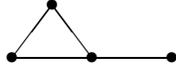
\begin{figure}[htb]
\setlength\unitlength{0.01\linewidth}
\begin{center}\begin{picture}(20,5)(-10,3)
\multiput(-12,0)(12,0)3{\makebox(0,0){$\bullet$}}
\put(-12,0){\line(1,0){24}}
\put(-6,8){\makebox(0,0){$\bullet$}}
\put(-12,0){\line(3,4){6}}
\put(0,0){\line(-3,4){6}}
\end{picture}\end{center}
\caption{A graph with different information complexity, maximum core size and
star packing rate}\label{fig:1}
\end{figure}

For trees, however, our lower and upper bounds coincide and we can even
compute this value efficiently.

\begin{theorem}\label{secondmain}
Let $G$ be a tree, let $c=c(G)$ be the maximum size of a core of $G$ and let
$s=s(G)$ be the star cover rate of $G$. For the information complexity
$\sigma(G)$ of $G$ we have
$$2-1/c=\sigma(G)= s.$$

One can compute $c$ and thus $\sigma(G)$ and the information rate $\rho(G)$ in
linear time.
Furthermore, a linear secret sharing scheme exists on $G$ that achieves
optimal load $2-1/c$. In this scheme the shares are vectors of length $2c-1$
over a finite field, the secret is a vector of length $c$ and these are
computed applying linear maps to a uniform random vector of some fixed length
less than $nc$, where $n$ is the number of vertices in $G$. The actual matrices
providing the linear maps can be found in time linear in the {\em output size}.
\end{theorem}

In Section \ref{sec:lower}
we prove the lower bound part of Theorem~\ref{thm:main} using the entropy
method, see \cite{capocelli,csirmaz1}. Note that the upper
bound comes from Stinson \cite{stinson}.

We prove the equalities of Theorem~\ref{secondmain} in section \ref{sec:upper}
by proving that $s(G)=2-1/c(G)$ if $G$ is a tree.

Finally in Section~\ref{sec:main} we prove the algorithmic assertions of
Theorem~\ref{secondmain}. 

\section{Information complexity of general graphs}\label{sec:lower}

In this section we show that the information complexity of an arbitrary graph 
is at least $2-1/c$ where $c$ is the size of the largest core in $G$. This 
proves the $2-1/c \le \sigma(G)$ part of Theorem \ref{thm:main}.

The proof uses the {\it entropy method}, see, e.g.~\cite{capocelli,
csirmaz1}. For the sake of completeness we sketch how this method works.
Consider any secret sharing scheme for an arbitrary access structure. For any
subset $A$ of the participants we define $f(A)$ to be the {\em normalized
entropy} of the shares belonging to the participants in $A$, namely
$$f(A)=\frac{H(\{S_v\mid v\in A\})}{H(\xi)},$$
where $S_v$ is the share of participant $v$ and $\xi$ is the secret. Note that
our goal is to lower bound the load of the scheme, which is
$\max_vf(\{v\})$.

Using the standard (Shanon-type) information inequalities we have
\begin{itemize}
\item[(a)] $f(\emptyset)=0$,
\item[(b)] $f(A)\ge f(B)$ when $A\supseteq B$ (monotonicity) and
\item[(c)] $f(A)+f(B)\ge f(A\cup B)+f(A\cap B)$ (submodularity).
\end{itemize}

Using the definition of the secret sharing schemes we further have 
\begin{itemize}
\item[(d)] $f(A)\ge f(B)+1$ when $A\supseteq B$, $A$ is qualified while $B$
is not (strict monotonicity) and
\item[(e)] $f(A)+f(B)\ge f(A\cup B)+f(A\cap B)+1$ when $A$, $B$ are qualified
while $A\cap B$ is not (strict submodularity).
\end{itemize}

The entropy method involves proving a lower bound for $\max_vf(\{v\})$ for any
$f$ satisfying inequalities (a)--(e). In our case we want to show that there
is always a vertex $v$ with $f(\{v\})\ge 2-1/c$ and this clearly follows from
the following lemma.

\begin{lemma}\label{thm:entropy}
Let $X$ be a core of the graph $G$, and let $f$ be a real valued
function defined on the subsets of the vertices of $G$ satisfying properties
{\rm (a)--(e)}. Then
$$
        \sum_{v\in X} f(\{v\}) \ge 2|X|-1.
$$
\end{lemma}
\begin{IEEEproof}
First observe that the statement is trivial if $|X|\le1$. We can therefore
assume $|X|\ge2$. We use  the ``independent sequence
lemma'' from \cite{blundo,csirmaz1} that ensures
$$
    f(X) \ge |X|+1.
$$
Using this inequality it is enough to prove
\begin{equation}\label{eq:firstthm}
    \sum_{v\in X} f(\{v\}) \ge f(X)+|X|-2 .
\end{equation}

We prove this latter inequality for all subsets $X$ that induce a connected
subgraph, not only for cores.
We use induction on the number of 
the vertices in $X$. The base case $X=\{v,w\}$ of (\ref{eq:firstthm})
simplifies to
$$
         f(\{v\})+f(\{w\}) \ge f(\{v,w\})
$$
which is subadditivity and a consequence of properties (a) and (c).

Now suppose $X$ induces a connected subgraph and it has at least three
vertices. Let us pick a
vertex $v\in X$ such that $Y=X-\{v\}$ also induces a connected subgraph. Note
that such a vertex
$v$ always exists. Let $w$ be a vertex in $Y$ connected to $v$. Neither
$\{v,w\}$ nor $Y$ is an independent set (we use $|X|\ge3$ here), but their
intersection $\{w\}$ is independent, thus unqualified. Property (e) gives
$$
    f(\{v,w\})+f(Y) \ge f(X)+f(\{w\})+1.
$$
Also, $f(\{v\})+f(\{w\})\ge f(\{v,w\})$ by subadditivity, which yields
$$
    f(\{v\})+f(Y) \ge f(X) +1.
$$
The induction hypothesis for $Y$ finishes the proof of (\ref{eq:firstthm}) and
also the proof of the lemma.
\end{IEEEproof}

\section{Information complexity of trees}\label{sec:upper}

In this section we show the equalities stated in
Theorem~\ref{secondmain}. They follow from Theorem~\ref{thm:main} and the
following lemma. To see this simply divide by $c$ the weights of the star
packing claimed by the lemma: the resulting fractional star packing shows that
star cover rate of $G$ is at most $2-1/c$.

\begin{lemma}\label{thm:3}
Let $G$ be a tree with at least 2 vertices, and suppose each core of $G$ has
size at most $c$. Then there exists a star packing in $G$
so that {\rm(i)} all edges are covered exactly $c$
times, and {\rm(ii)} all vertices are covered at most $2c-1$ times.
\end{lemma}

\begin{IEEEproof}
We replace each undirected edge $(u,v)$ of $G$ by $c$ directed edges
between $u$ and $v$; the number of edges in each direction will be specified
later.

To obtain the star packing we partition the (now directed) edges into stars in
such a way that all edges will be directed outward from the center of the
star. Thus all outgoing edges from a vertex $v$ must be part of stars centered
at $v$. Clearly, we can do this with as many stars centered at $v$ as the
maximal number of {\em outgoing} edges from $v$ to some neighboring
vertex. Furthermore $v$ will be a non-center vertex of exactly as many stars
as the total number of {\em incoming} directed edges to $v$.
The sum of these two numbers gives the total number of stars covering $v$.
As there are exactly $c$ directed edges along each original edge, this cover 
number is $c$ 
plus the total number of incoming directed edges {\em except the 
smallest number} of incoming directed edges from a single neighbor.

Thus it suffices to show that we can direct these multiple
edges so that this latter sum is at most $c-1$.

We start with assigning positive integers -- weights -- to each vertex. 
The weight of a set of vertices is the sum of the weights of the vertices in
the set. Assigning weights is a technical step to ensure each vertex is in a
maximum weight core.

Let $\mathcal W$ be the set of all positive integer weight functions making
the weight of every core at most $c$. As each vertex is an element of some core, $\mathcal W$ has
finitely many elements. Furthermore $\mathcal W$ is not empty: if every
vertex has weight $1$, then by the definition of $c$, every core has weight
$\le c$.
We call a weight function $\w\in\mathcal W$ {\em maximal} if increasing $\w$ by
one at any one vertex yields a function outside $\mathcal W$. Clearly, a
maximal weight function must exist in $\mathcal W$. 

From now on fix such a maximal weight function $\w\in\mathcal W$. The
maximality of $w$ implies that for every vertex $v$ there exists a core
containing $v$ whose weight is exactly $c$.

\begin{figure}[htb]
\setlength\unitlength{0.007\linewidth}
\begin{center}\begin{picture}(90,33)(0,-8)
\put(0,0){\line(1,0){18}}
\multiput(22,0)(10,0)6{\line(1,0){6}}
\put(82,0){\line(1,0){8}}
\put(90,0){\makebox(0,0){$\bullet$}}
\put(0,0){\makebox(0,0){$\bullet$}}
\put(10,0){\makebox(0,0){$\bullet$}}
\put(19.6,-0.3){\makebox(0,0){$A$}}
\put(29.8,-0.3){\makebox(0,0){$B$}}
\put(39.8,-0.3){\makebox(0,0){$C$}}
\put(49.8,-0.3){\makebox(0,0){$D$}}
\put(59.7,-0.3){\makebox(0,0){$E$}}
\put(69.7,-0.3){\makebox(0,0){$F$}}
\put(79.8,-0.3){\makebox(0,0){$G$}}
\put(20,-2){\line(0,-1){8}}
\put(30,-2){\line(0,-1){8}}
\put(60,-2){\line(0,-1){8}}
\put(70,-2){\line(0,-1){8}}
\put(20,2){\line(0,1){8}}
\put(30,2){\line(0,1){8}}
\put(60,2){\line(0,1){8}}
\put(70,2){\line(0,1){8}}
\put(20,10){\makebox(0,0){$\bullet$}}
\put(30,10){\makebox(0,0){$\bullet$}}
\put(60,10){\makebox(0,0){$\bullet$}}
\put(70,10){\makebox(0,0){$\bullet$}}
\put(20,-10){\makebox(0,0){$\bullet$}}
\put(30,-10){\makebox(0,0){$\bullet$}}
\put(60,-10){\makebox(0,0){$\bullet$}}
\put(70,-10){\makebox(0,0){$\bullet$}}
\put(20,10){\line(-1,1){7}}
\put(30,10){\line(-1,1){7}}
\put(30,10){\line(1,1){14}}
\put(60,10){\line(1,1){7}}
\put(70,10){\line(1,1){7}}
\put(13,17){\makebox(0,0){$\bullet$}}
\put(23,17){\makebox(0,0){$\bullet$}}
\put(37,17){\makebox(0,0){$\bullet$}}
\put(44,24){\makebox(0,0){$\bullet$}}
\put(67,17){\makebox(0,0){$\bullet$}}
\put(77,17){\makebox(0,0){$\bullet$}}
\put(1.5,-3){\makebox(0,0){$\scriptstyle7$}}
\put(91.5,-3){\makebox(0,0){$\scriptstyle7$}}
\multiput(15.5,19)(10,0)2{\makebox(0,0){$\scriptstyle7$}}
\multiput(69.5,19)(10,0)2{\makebox(0,0){$\scriptstyle7$}}
\multiput(21.5,-11.5)(10,0)2{\makebox(0,0){$\scriptstyle7$}}
\multiput(61.5,-11.5)(10,0)2{\makebox(0,0){$\scriptstyle7$}}
\put(45.5,22){\makebox(0,0){$\scriptstyle7$}}
\multiput(11.5,-3)(10,0){4}{\makebox(0,0){$\scriptstyle1$}}
\put(51.5,-3){\makebox(0,0){$\scriptstyle2$}}
\multiput(61.5,-3)(10,0)3{\makebox(0,0){$\scriptstyle1$}}
\multiput(21.5,8)(10,0)2{\makebox(0,0){$\scriptstyle1$}}
\multiput(61.5,8)(10,0)2{\makebox(0,0){$\scriptstyle1$}}
\put(38.5,15){\makebox(0,0){$\scriptstyle1$}}
\end{picture}\end{center}
\caption{A tree with weights and maximal core size $c=7$.}\label{fig:2}
\end{figure}
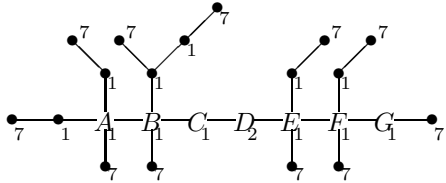

Now let $(v_1,v_2)$ be an edge of $G$. If either $v_1$ of $v_2$ is a leaf, then 
direct all $c$ edges between $v_1$ and $v_2$ toward the leaf. 
(If both $v_1$ and $v_2$ are
leaves, then $G$ is a single edge, and there is nothing to prove.) 

If neither $v_1$ nor $v_2$ is a leaf, then removing the edge $(v_1,v_2)$
splits $G$ into two disjoint subtrees, $G_1$ and $G_2$ where $G_i$ 
contains $v_i$. Let $C_i$ be a maximal weight (using the weight function
$\w$) core in $G_i$ such that
$C_i$ contains $v_i$ and let its weight be $c_i=\w(C_i)$.
As $C_1\cup C_2$ is a core of weight $c_1+c_2$ in $G$, and all cores in $G$
has weight $\le c$, we have 
$c_1+c_2 \le c$. Among the $c$ directed edges between $v_1$ and $v_2$ 
direct $c_1$
from $v_1$ towards $v_2$, and $c_2$ from $v_2$ towards $v_1$. If $c_1+c_2<c$
then direct the rest of these edges arbitrarily.

The tree depicted on figure \ref{fig:2} has maximal core size $c=7$, and 
the numbers show a maximal weight function. Each edge is replaced by seven
directed edges, and the numbers the above procedure gives are
\begin{center}
\begin{tabular}{cc@{~~}cc@{~~}cc}
$A\rightarrow B$ & $B\rightarrow C$
  & $C\rightarrow D$ & $D\rightarrow E$ & $E\rightarrow F$ & $F\rightarrow G$ \\
$\scriptstyle3$ & $\scriptstyle6$ & $\scriptstyle{}\ge1$ & $\scriptstyle2$ &
$\scriptstyle4$ & $\scriptstyle6$ \\[5pt]
 $A\leftarrow B$ & $B\leftarrow C$ & $C\leftarrow D$ & $D\leftarrow E$ & $E\leftarrow F$
  & $F\leftarrow G$ \\
$\scriptstyle4$ & $\scriptstyle1$ & $\scriptstyle{}\ge2$ & $\scriptstyle5$ &
$\scriptstyle3$ & $\scriptstyle1$
\end{tabular}
\end{center}
For example, when the edge $CD$ is deleted, the only core in the remaining
graph containing $D$ is the singleton $\{D\}$ with weight $2$. This gives the
value $\ge2$ to 
$C\leftarrow D$ and similarly we have $\ge 1$ for $C\rightarrow D$. This
leaves $4$ more edges between $C$ and $D$ that we can direct arbitrarily. In all
other edges in the above example we have $c_1+c_2=c$, thus the direction of
all other edges are determined.

We claim that our construction satisfies the above requirement. Indeed, if $v$
is a leaf, then it has exactly $c$ incoming edges and no outgoing edge.
Otherwise let $v$ be a non-leaf vertex, and $C$ be a core of maximal weight
(according to $\w$) containing $v$. By the maximality of $\w$, $C$ has 
weight $c$. When deleting $v$ from $C$ each
connected component of the remaining graph contains exactly one neighbor of
$v$ in $C$.  Let $v_1$, $v_2$, $\dots$,
$v_s$ be these neighbors and let $C_i$ be the connected component of $C-v$
containing $v_i$. Then
$$
    c= \w(C) = \w(v)+\w(C_1) + \cdots + \w(C_s) .
$$
Both $C$ and $C-C_i$ are cores in $G-vv_i$ and they were considered when directing the edges along
$vv_i$. Therefore we have at least $\w(C_i)$ edges directed from $v_i$ to $v$
and at least $\w(C-C_i)=c-\w(C_i)$ edges going from
$v$ to $v_i$. As this accounts for all $c$ edges between $v$ and $v_i$ these
are the exact number of edges going either way. Thus the total number of {\em
incoming} edges to $v$ from vertices in $C$ is
$$
   \w(C_1) + \dots + \w(C_s) = c - \w(v) \le c-1 .
$$
We have two cases: either $v$ has a leaf neighbor, or it has none. In the
first case all non-leaf neighbors of $v$ are in $C$, as $C$ was chosen
to be maximal. There are no incoming edges from leaves, thus in this case
we are done.

In the other case no neighbor of $v$ is a leaf. Again by maximality all but
one of the neighbors of $v$ must 
be in $C$. Let $v^*$ be the exceptional neighbor of $v$ outside $C$. 
Now $C-C_i$ is a core in the graph $G-vv^*$ and it contains $v$,
thus at least $\w(C-C_i)=c-\w(C_i)$ edges are directed from $v$ toward 
$v^*$. It means that that the number of {\em incoming} edges from $v^*$
cannot be more than $\w(C_i)$, which is the number of incoming edges from $v_i$.
It shows that the smallest number of incoming edges come from $v^*$, and
the total number of incoming edges from the other neighbors is at most
$c-1$, which was to be shown.
\end{IEEEproof}

\section{Algorithms}\label{sec:main}

We turn to the algorithmic part of Theorem~\ref{secondmain}. Let
$G$ be a tree.
The size $c(G)$ of the maximal core in $G$ can be found by the following
algorithm.

Pick an arbitrary root $r$ in $G$. For each
vertex $v$ in $G$ let us denote by $G_v$ the subtree of $G$ ``below''
$v$, i.e., $G_r=G$ and for $v\ne r$ we obtain $G_v$ by deleting the edge
connecting $v$ to its ``parent'' (the neighbor closer to $r$) and taking the
connected component of $v$.

First we order the the vertices in reverse breadth first search order
(starting from the vertices farthest from the root) and compute the value
$c(v)$ of the size of the largest core in $G_v$ containing $v$. We define
$c(v)=0$ for leaf vertices $v$. If $v$ is not a leaf, then $c(v)$ is one plus
the sum of $c(v_i)$ for all children $v_i$ of $v$ {\em with the smallest
summand left out of the summation}. This enables us to compute $c(v)$ in time
$O(d_v)$ from the values computed earlier. Here $d_v$ stands for the degree of
$v$. This makes for a linear time algorithm for computing all the values
$c(v)$.

Having computed $c(v)$ for each vertex, computing $c(G)$ is simple. If the
largest core contains the root $r$, then its size is $c(r)$. Otherwise if
$v\ne r$ is its vertex closest to the root its size is one plus the sum of
$c(v_i)$ for all the children $v_i$ of $v$ (this time no summand is left
out). Computing these values and finding the maximum takes linear time again.

Finally in order to construct the optimal secret sharing scheme one has to
find a maximal weight function $\w\in\mathcal W$. Notice that for an arbitrary
weight function $\w$ one can compute all the values $c_\w(v)$ in linear time
the same way we computed $c(v)$. Here $c_\w(v)$ is the maximal $\w$-weight of a
core in $G_v$ containing $v$. Now increasing the weight of the root $r$ by
$c-c_\w(r)$ we can ensure that no core has weight over $c$ but the root is
contained in a core of weight $c$. Starting from the all $1$ weight function and
repeating this procedure for all vertices as roots we find a maximal weight
function. This takes quadratic time (still OK as the output is huge), but we
remark that with a more careful
analysis (increasing the weight of vertices in a single breadth first search
order after computing first $c(v)$ without weights) a maximal weight function
can be also obtained in linear time.

From a maximal weight function $w$ one can orient $c_w(v)$ edges from $v$ to
its parent ($v\ne r$) and $c-c_w(v)$ edges from the parent to $v$. This yields
an optimal star packing. Now we apply Stinson's technique \cite{stinson} to
obtain the secret sharing scheme on $G$ by combining linear schemes on the
individual stars. The parameters of this combined scheme are as stated in
Theorem~\ref{secondmain}.

\begin{IEEEbiographynophoto}{L\'aszl\'o Csirmaz}
has been with Central European University, Budapest, since 1996. Before that
he worked as a researcher at the R\'enyi Institute of Mathematics, Budapest.
His main research interests include secret sharing, Shannon theory, and
combinatorial games.
\end{IEEEbiographynophoto}

\begin{IEEEbiographynophoto}{G\'abor Tardos}
received his PhD in mathematics from E\"otv\"os University in 1988.
A research fellow at the Re\'enyi Institute, Budapest, Hungary since 1990.
Canada Research Chair of Computational and Discrete Geometry at the Simon
Fraser
University, BC, Canada, since 2005.

His main research interests are combinatorics, discrete and computational
geometry, and complexity theory.
\end{IEEEbiographynophoto}

\vfill

\end{document}